\documentstyle[aps]{revtex}\tighten

\def \D {\mbox{D}}
\def \rd {\displaystyle{\cdot}}

\begin{document}

\title{Magnetized cosmological perturbations}
\author{Christos G. Tsagas\thanks{christos.tsagas@port.ac.uk}
and Roy Maartens\thanks{roy.maartens@port.ac.uk}}
\address{Relativity and Cosmology Group,
Division of Mathematics and Statistics, Portsmouth University,
Portsmouth~PO1~2EG, England}
\maketitle

\begin{abstract}

A large-scale cosmic magnetic field affects not only the growth of
density perturbations, but also rotational instabilities and
anisotropic deformation in the density distribution. We give a
fully relativistic treatment of all these effects, incorporating
the magneto-curvature coupling that arises in a relativistic
approach. We show that this coupling produces a small enhancement
of the growing mode on superhorizon scales. The magnetic field
generates new nonadiabatic constant and decaying modes, as well as
nonadiabatic corrections to the standard growing and decaying
modes. Magnetized isocurvature perturbations are purely decaying
on superhorizon scales. On subhorizon scales before recombination,
magnetized density perturbations propagate as magneto-sonic waves,
leading to a small decrease in the spacing of acoustic peaks.
Fluctuations in the field direction induce scale-dependent
vorticity, and generate precession in the rotational vector. On
small scales, magnetized density vortices propagate as Alfv\'{e}n
waves during the radiation era. After recombination, they decay
slower than non-magnetized vortices. Magnetic fluctuations are
also an active source of anisotropic distortion in the density
distribution. We derive the evolution equations for this
distortion, and find a particular growing solution.

\end{abstract}

\pacs{98.80.Hw, 04.40.Nr, 47.75.+f, 95.30.Qd, 98.62.En}

%%%%%%%%%%%%%%%%%%%%%%
\section{Introduction}
%%%%%%%%%%%%%%%%%%%%%%

Recent observations reveal the widespread
existence of magnetic fields in the universe and are producing
much firmer estimates of their strengths in interstellar and
intergalactic space. They also appear to be a common property of
the intracluster medium of galaxy clusters, extending well
beyond the core regions (see \cite{K} and references
therein). Strengths of ordered magnetic fields in the intracluster
medium of cooling flow clusters exceed those typically associated
with the interstellar medium of the Milky Way, suggesting that
galaxy formation and even cluster dynamics are, at least in some
cases, influenced by magnetic forces. Furthermore, reports of
Faraday rotation associated with high redshift Lyman-$\alpha$
absorption systems seem to imply that dynamically significant
magnetic fields may be present in condensations at high redshift
\cite{KPZ}. The more we look for
extragalactic magnetic fields, the more ubiquitous we find them
to be.

Large-scale magnetic fields introduce new ingredients into the
standard, but nevertheless uncertain, picture of the early
universe. They seem unlikely to survive an epoch of inflation,
but it is conceivable that large-scale fields and magnetic
inhomogeneities could be generated at the end of that era
or in subsequent phase transitions (see, e.g., \cite{R}).
Studies of magnetogenesis are partly
motivated by the need to explain the origin of large-scale galactic
fields. Typical spiral galaxies have magnetic fields of the order
of a few $\mu$G coherent over the plane of their disc. Such fields
could arise from a relatively large primordial seed field,
amplified by the collapse of the protogalaxy, or by a much weaker
one that has been strengthened by the galactic dynamo. Provided that
this mechanism is efficient, the seed can be as low as
$\sim10^{-30}$G
at present \cite{dlt}. However, in the absence of nonlinear amplification,
seeds of the order of $10^{-12}$G or even $10^{-8}$G are required
\cite{Ku}.

Determining whether the origin of galactic and cluster magnetic
fields is primordial or post-recombination is a difficult task,
since strong amplification in these virialized systems
overwhelms any traces of their earlier history. In contrast,
magnetic effects on the cosmic microwave background (CMB) anisotropies,
or any magnetic presence away from clusters and galaxies, can provide better
insight into these early phases. If large-scale magnetic fields
are present throughout the universe today, their structure and
spectrum should bear clearer signatures of their past. Thus,
improved direct observations, such as high resolution Faraday
rotation maps and the study of extragalactic cosmic rays, may
help in this respect \cite{O}. For example, we would like to know
whether or not the intergalactic voids are permeated by a widespread
magnetic field, and whether there is magnetic
field evolution in galaxies. If large-scale magnetic fields were
present in the early universe, were they dynamically significant,
and if so, how have they affected the formation and evolution of
the observed structure? It is known that element abundances
constrain the strength of a primordial field at
the nucleosynthesis epoch \cite{gr}. Stronger
limits on a primordial magnetic field are imposed via
CMB anisotropies, since the field distorts the
acoustic peaks and induces Faraday rotation in
the polarization \cite{adgr,cmb,dky}.

In this article we assume the existence of a large-scale
ordered magnetic field of primordial origin  {\em a priori}, and we
investigate the magnetic effects on density inhomogeneities.
Specifically, we
analyze magnetized density perturbations, magnetized
cosmic vortices (i.e., rotational instabilities), and
magnetized shape distortion.
Magnetized density perturbations were studied by Ruzmaikina and
Ruzmaikin \cite{RR} in Newtonian theory, while Wasserman \cite{W}
looked at the rotational behavior of a magnetized fluid.
Kim et al. \cite{kor} derive a magnetized Jeans length, assuming
that there are no density perturbations in the absence of the
field.
In a relativistic treatment, Battaner et al. \cite{bfj}
investigated magnetized
structure formation in the radiation era. Jedamzik et al. and
Subramanian \& Barrow
\cite{sb} have considered magnetic dissipative effects
at recombination.

We generalize aspects of these previous treatments by giving a
fully relativistic analysis of the scalar and vector contributions
of the magnetic field to the evolution of density
inhomogeneity. We
consider not only density perturbations and rotational
instabilities, but also the shape-distortion effects of the field.
Density perturbations are found explicitly in the radiation and
dust eras, including a new solution that shows how the
relativistic magneto-curvature coupling acts to enhance growth on
superhorizon scales. The existence of the magneto-curvature
coupling was first identified by Tsagas
and Barrow \cite{TB}.
The nonadiabatic magnetic effect on the modes
is clearly identified, including the magnetized isocurvature
modes. New solutions are also found for rotational
instabilities, which are significantly affected by the field,
and we show that magnetic effects actively generate shape distortion in the
density distribution.
We follow the relativistic analysis of cosmic electromagnetic fields
given by Ellis \cite{E2}, and we use the covariant and gauge-invariant
approach to perturbations \cite{EB,EB2,MT}. A covariant and
gauge-invariant analysis of magnetized density perturbations was
first developed by Tsagas and Barrow \cite{TB,T}, whose results we
extend.

We adopt the usual approximation that in the background,
which is a spatially flat Friedmann-Robertson-Walker (FRW)
model,\footnote{
Spatial flatness is necessary for the gauge-invariance of
all the perturbative variables \cite{TB}.}
the ordered large-scale magnetic field is too weak to destroy
spatial isotropy.
The weak field approximation is an acceptable physical approximation when the
field energy density is a small fraction of the isotropically distributed
dominant energy density (see \cite{weak} for further discussion).
If one demands strict mathematical homogeneity in the background,
i.e., if one refuses to accept a coherent test field in the background,
then one must adopt a Bianchi I model for the background. However,
this is a highly complicated approach, which in the end will
give results that are practically indistinguishable from those
with an FRW background. (We are currently completing calculations
that confirm this statement \cite{tm}.)
The standard assumption of very high
conductivity is also made, so that we can ignore
large-scale electric fields, while maintaining the desired
coupling between the fluid and the magnetic field.
We use a single perfect fluid model, which is reasonable in the radiation era,
but does not apply during recombination, while
after last scattering, it means that our solutions only apply to a
baryon-dominated universe. (See \cite{ck} for a discussion of the effects of
cold dark matter (CDM) potential wells on the field.)

The assumption of a weak background magnetic field ensures that the
field terms in our linearized equations are first-order. Current
limits on the strength of a primordial magnetic field show that its
influence is secondary relative to that of the dominant matter
component. In practice, this allows for the possibility that
second-order fluid terms can have a strength comparable to that of
the linear magnetic terms. Nevertheless, such second-order terms do
not contain any further information regarding the lowest order
influence of the field on gravitational instability. We ignore the
second-order fluid terms, even though they may be of comparable
magnitude to the first-order magnetic effects. This approach is
consistent at linear order, and allows us to
isolate the lowest order magnetic effects
on gravitational instability.
Our aim is to identify the sources of the magnetic
effects, calculate their impact to lowest order, and discuss
their implications for the evolution of density inhomogeneities.

In Sec. II, we outline the formalism and the main equations that
govern the coupled evolution of density inhomogeneity, the magnetic field
and the curvature.
Sec. III considers magnetized density perturbations, identifying the
nonadiabatic effects of the field.
We find a new solution on superhorizon scales in the radiation
era, showing how the magneto-curvature coupling slightly enhances
growth. In the radiation era, a
small damping effect is wrongly predicted when the magneto-curvature
coupling is ignored. On subhorizon scales, magneto-sonic waves in
the radiation era have a slightly increased frequency, leading to
a decrease in the spacing of CMB acoustic peaks. In the dust era,
the growing mode on small scales is slightly damped by magnetic
effects.
We also find the pure-magnetic density perturbations,
i.e., the fluctuations created in a smooth plasma at
magnetogenesis. These include growing modes.
Magnetized isocurvature modes are characterized, and
found explicitly on superhorizon scales. These modes all decay,
in the radiation and dust eras.
Magnetized cosmic vortices are considered in Sec. IV. We show that
magnetized vorticity is scale-dependent , and that the field
generates
precession in the rotational vector. We solve exactly for the
rotational instabilities, showing that they propagate as
Alfv\'{e}n (vector) waves on small scales
during the radiation era. After recombination, such vortices
persist for longer than non-magnetized vortices.
Sec. V investigates magnetized
shape distortion, showing that the field is as an active
source of distortion. Purely magnetic distortion on superhorizon scales in the
dust era is shown to have a growing mode.
Conclusions are given in Sec. VI.

We use units with $c=1=8\pi G$ and our signature is $(-+++)$;
spacetime indices are $a,b,\cdots$ and (square) round brackets
enclosing indices denote (anti-)symmetrization.

%%%%%%%%%%%%%%%%%%%%%%%%%%%%%%%%%%%%%
\section{Cosmic magnetohydrodynamics}
%%%%%%%%%%%%%%%%%%%%%%%%%%%%%%%%%%%%%

As noted above, the cosmic magnetic field $B^a$ must have weak energy
density $\rho_{\rm mag}={1\over2}B_aB^a$
to be consistent with observational limits, so that
$c_{\rm a}\ll1$,
where $c_{\rm a}$ is the Alfv\'{e}n speed. The
Alfv\'{e}n speed, which effectively leads to a nonadiabatic increase in the sound
speed, is given by
\begin{equation}
c_{\rm a}^2= {B^2\over\rho} \,, \label{1}
\end{equation}
where $\rho$ is the energy density of the cosmic fluid.
In the limit of vanishing density inhomogeneity, i.e., in the
 background, the field is uniform, but its weak magnitude means that it
does not
disturb the background isotropy, so that the magnetic anisotropic
stress is negligible in the background.
In the actual inhomogeneous universe the
field's influence propagates via:
%\begin{itemize}
%\item
%the magnetic energy density and pressure
%($p_{\rm mag}={1\over3}\rho_{\rm mag}$) and their associated
%spatial gradients, which affect all three types of density
%inhomogeneity;
%\item
%fluctuations in the direction of the magnetic vector itself,
%coupled to the orientation of the background field, with an impact
%on rotational instabilities;
%\item
%anisotropies in the distribution of the field vector distorting
%the shape of density condensations.
%\end{itemize}
\begin{enumerate}
\item[]
the background energy density and pressure
($p_{\rm mag}={1\over3}\rho_{\rm mag}$) occur in terms of the form
$c_{\rm a}^2{\cal P}$, where ${\cal P}$ is a perturbed quantity, and
despite their weakness, they
can have observable consequences (e.g., a change in the spacing of
CMB acoustic peaks in the radiation era);
\item[]
spatial gradients of $\rho_{\rm mag}$ couple with gradients of $\rho$
and thus alter the fluctuations of $\rho$ (in particular,
introducing nonadiabatic modes);
\item[]
the background direction of the field introduces anisotropy by
picking out preferred directions in perturbed vector and tensor fields,
and preferred directional derivatives of perturbed scalar/ vector/ tensor fields,
leading to effects such as Faraday rotation;
\item[]
the background direction of the field is also the source of the
magneto-curvature coupling, via
terms of the form $K_{abcd}B^d$, where $K_{abcd}$ is the part of
the curvature tensor which vanishes in the background;
\item[]
fluctuations in the direction of the field
 generate new anisotropies that can source magnetized
vortices (leading in particular to Alfv\'{e}n waves) and shape distortion.
\end{enumerate}

We include all of these aspects in our analysis, so that we
incorporate the full range of scalar
(magnetic energy density and
isotropic pressure) and vector (anisotropic stress) effects of the
field, allowing for fluctuations in both the magnitude and direction.
In order to provide a transparent relativistic generalization
of Newtonian analysis, and to use
variables that as far as possible have a direct physical
interpretation,
we adopt a covariant Lagrangian approach \cite{E2,EB,EB2,MT,M1,mb}.
This continues and develops the work of \cite{TB}. In particular, we
discuss in detail the physical meaning and implications of the
density perturbation solutions, and we extend the investigation to
cover magnetized cosmic vortices and shape distortion.

The cosmic perfect fluid defines a unique
four-velocity $u_a$ (with $u_au^a=-1$), and then
$h_{ab}=g_{ab}+u_au_b$,
where $g_{ab}$ is the spacetime metric, projects
into the local rest spaces of comoving
observers. The projection of a vector is
$V_{\langle a\rangle}=h_a{}^bV_b$, and
a projected second rank tensor
$S_{ab}$ splits irreducibly as
\[
S_{ab}={\textstyle{1\over3}}Sh_{ab}
+\varepsilon_{abc}S^c+
S_{\langle ab\rangle }\,,
\]
where
$S\equiv h_{ab}S^{ab}$ is the spatial trace,
$S_a\equiv{1\over2}\varepsilon_{abc}S^{bc}$ is the spatial
vector dual to the skew part of $S_{ab}$, and
$S_{\langle ab\rangle}\equiv[h_{(a}{}^ch_{b)}{}^d-{1\over3}h^{cd}
h_{ab}]S_{cd}$ is the projected symmetric tracefree (PSTF) part.
 Here
$\varepsilon_{abc}=\eta_{abcd}u^d$ is the
projection of $\eta_{abcd}$, the spacetime alternating
tensor.

The covariant derivative splits into a comoving
time derivative $\dot{J}_{a\cdots b}=u^c\nabla_cJ_{a\cdots b}$,
and a covariant spatial derivative
$\D_cJ_{a\cdots b}=h_c{}^dh_a{}^e\cdots h_b{}^f\nabla_dJ_{e\cdots f}$.
Then we define a covariant spatial divergence and curl that generalize
the Newtonian operators to curved spacetime \cite{M1}:
\begin{eqnarray*}
 \mbox{div}\,V= \D^aV_a\,,&~~& (\mbox{div}\,S)_a= \D^bS_{ab}\,,\\
\mbox{curl}\,V_a=\varepsilon_{abc}\D^bV^c\,,&~~&
\mbox{curl}\,S_{ab}=\varepsilon_{cd(a}\D^cS_{b)}{}^d\,.
\end{eqnarray*}

The fluid kinematics are described by the expansion
$\Theta=\mbox{div}\,u$, four-acceleration $A_a=\dot{u}_a$,
vorticity $\omega_a=-{1\over2}\mbox{curl}\,u_a$ and shear
$\sigma_{ab}=\D_{\langle  a}u_{b\rangle}$.
Local curvature is described by the Ricci tensor $R_{ab}$, while
nonlocal tidal forces and gravitational radiation are described by
the electric and the magnetic parts of the Weyl tensor,
$E_{ab}=E_{\langle ab\rangle }= C_{acbd}u^cu^d$
and $H_{ab}=H_{\langle ab\rangle }={1\over2}
\varepsilon_{acd}C^{cd}{}{}_{be}u^e$.
The magnetized perfectly
conducting fluid has energy density
$\rho$ and isotropic pressure $p$.
The magnetic field is $B_a=B_{\langle a\rangle}$,
with energy density, isotropic
pressure and anisotropic stress given respectively by
\begin{equation}
\rho_{\rm mag}={\textstyle{1\over2}}B_aB^a\,,~p_{\rm mag}=
{\textstyle{1\over3}}\rho_{\rm mag}\,,~
\pi_{ab}=-B_{\langle a}B_{b\rangle}\,.
\label{emt'}\end{equation}
Then the total energy-momentum tensor is
\begin{equation}
T_{ab}=\left(\rho+\rho_{\rm mag}\right)u_au_b+ \left(p+p_{\rm
mag}\right)h_{ab}+\pi_{ab} \,.  \label{emt}
\end{equation}
Notice that the absence of an electric
field means that there is no energy flux (Poynting vector).
The magnetic field appears from Eq. (\ref{emt}) to behave like
a radiation fluid with anisotropic stress. However, this fluid
picture does not fully encompass the vector properties of the
field, and in
particular, its coupling to the curvature.

In the background, $B_a$ is weak enough not to affect the
isotropy, i.e. the anisotropic stress is negligible in
the background, and $\rho_{\rm mag}\ll \rho$.
The background expansion is $\Theta=3H$, where $H=\dot{a}/a$ is
the Hubble rate.
The background is
covariantly characterized by
\begin{eqnarray}
&&\D_a\Theta=\D_a\rho=\D_ap=\D_bB_a=0 \,,  \label{nograd}\\
&&A_a=\omega_a=0 \,,  \label{novec}\\
&&\sigma_{ab}=E_{ab}=H_{ab}={\cal R}_{ab}=\pi_{ab}=0 \,,
\label{noten}
\end{eqnarray}
where
\[
{\cal R}_{ab}=
h_a{}^{c}h_b{}^{d}R_{cd}+R_{acbd}u^cu^d+\D_cu_a\D_bu^c-\Theta\D_bu_a\,,
\]
with $R_{ab}$ the Ricci tensor
and $R_{abcd}$ the Riemann tensor.
Note that ${\cal R}_{ab}$ is the intrinsic 3-Ricci tensor
of spatial hypersurfaces only if $\omega_a=0$; otherwise there
are no such hypersurfaces orthogonal to $u^a$ \cite{E2}.

Quantities that vanish in the background
are gauge-invariant, and they covariantly describe linear deviations
from homogeneity and anisotropy.
We collect below the linearized evolution and constraint equations
given in \cite{TB}, rewritten in the streamlined formalism of
\cite{M1}, which considerably simplifies the equations and facilitates
analysis of their properties.
The following covariant identities \cite{MT} are used in deriving the
equations (assuming a flat background
with vanishing cosmological constant):
\begin{eqnarray}
{\rm curl}\,\D_af &=&-2\dot{f}\omega_a \,,
\label{id2} \\
\left(a\D_af\right)^{\rd} &=& a\D_a\dot{f}+a\dot{f}A_a \,,
\label{a14}\\
\D^2\left(\D_af\right) &=&\D_a\left(\D^2f\right)
+2\dot{f}{\rm curl}\,\omega_a
\,, \label{a19}\\
\left(a\D_aJ_{b\cdots}\right)^{\rd} &=& a\D_a\dot{J}_{b\cdots}\,,
\label{id1}\\
\D_{[a}\D_{b]}V_c &=&0=
\D_{[a}\D_{b]}S^{cd}
\,, \label{a17}\\
{\rm div}\,{\rm curl}\, V &=& 0 \,,\label{a20}\\
({\rm div}\,{\rm curl}\, S)_{a} &=& {\textstyle{1\over2}}{\rm curl}\, ({\rm div}\,
 S)_{a}\,,\label{a18}\\
{\rm curl}\,{\rm curl}\, V_a &=& \D_a ({\rm div}\, V)
-\D^2V_a\,,
\label{id3}\\
{\rm curl}\,{\rm curl}\, S_{ab} &=& {\textstyle{3\over2}}\D_{\langle a}({\rm div}\,
 S)_{b\rangle}-\D^2S_{ab}\,,\label{a23}
\end{eqnarray}
where the vectors and tensors vanish in the background,
$S_{ab}=S_{\langle ab\rangle}$. The magnetic field itself does not
vanish in the background, so that its projected derivatives do not
commute at linear order: the vector identity in Eq. (\ref{a17}) is replaced by
\[
\D_{[a}\D_{b]}B_c={\textstyle{1\over2}}{\cal
R}_{dcba}B^d-\varepsilon_{abd}\omega^d\dot{B}_c\,,
\]
where ${\cal R}_{abcd}$ is formed from $R_{abcd}$ and the
kinematic quantities \cite{TB}. This non-commutativity is the root
of the magneto-curvature coupling found in \cite{TB}.

%%%%%%%%%%%%%%%%%%%%%%%%%%%%%%%%
\subsection{Maxwell's equations}
%%%%%%%%%%%%%%%%%%%%%%%%%%%%%%%%

In the infinite
conductivity limit, Maxwell's equations \cite{E2,mb} provide three constraints,
\begin{eqnarray}
\mbox{div}\,B&=&0  \,,\label{Max1}\\
\mbox{curl}\,B_a&=&\varepsilon_{abc}B^bA^c+j_a \,,\label{m'}\\
\omega^aB_a &=& {\textstyle{1\over2}}q\,, \label{m''}
\end{eqnarray}
where $j_a$ is the current and $q$ the charge density
generated by fluctuations,
and one propagation equation
\begin{equation}
\dot{B}_{\langle  a\rangle}=-{\textstyle{2\over3}}\Theta B_a+
\sigma_{ab}B^b+ \varepsilon_{abc}B^b\omega^c \,,  \label{Max2}
\end{equation}
which is the covariant form of the induction equation.
Note that
$\dot{B}_{\langle  a\rangle}=\dot{B}_a-A_bB^bu_a$,
and $B^aA_a=0$ to first order
 only in the case of a pressure-free perfect fluid \cite{TB}.

Contracting Eq. (\ref{Max2}) with $B^a$, and neglecting the second order term
$\sigma_{ab}\pi^{ab}$, we deduce
the radiation-like evolution law of the magnetic
energy density,
\begin{equation}
(B^2)^{\rd}+{\textstyle{4\over3}}\Theta(B^2)=0
\,.  \label{B^2ev}
\end{equation}
We can also derive the evolution of the anisotropic stress
from Eq. (\ref{Max2}):
\begin{equation}
\dot{\pi}_{ab}=-4H\pi_{ab}-{\textstyle{2\over3}}c_{\rm
a}^2\rho\sigma_{ab}\,,\label{pi}
\end{equation}

%%%%%%%%%%%%%%%%%%%%%%%%%%%%%%
\subsection{Conservation laws}
%%%%%%%%%%%%%%%%%%%%%%%%%%%%%%

Energy density conservation is expressed via the equation of
continuity
\[%\begin{equation}
\dot{\rho}+\Theta(1+w)\rho=0 \,,
\]%\label{edc}\end{equation}
where $w=p/\rho$.
Notice the absence of magnetic
terms in this equation, since field energy conservation holds
separately as a consequence of Maxwell's equations, as shown in
Eq. (\ref{B^2ev}). The two energy conservation equations imply
\[
\left(c_{\rm a}^2\right)^{\rd}=(w-{\textstyle{1\over3}})\Theta
c_{\rm a}^2\,.
\]

On the other hand, the field does enter conservation of
momentum density
\begin{equation}
(1+w)\rho A_a+c_{\rm s}^2\D_a\rho+
\varepsilon_{abc}B^b\mbox{curl}\,B^c=0 \,,  \label{mdc}
\end{equation}
where
$c_{\rm s}^2=\dot{p}/\dot{\rho}$ is the adiabatic sound-speed
squared (with $\D_ap=c_{\rm s}^2\D_a\rho$).
This equation reflects the momentum density exchange between the
fluid and the field.
The magnetic field is a source of acceleration (provided
that curl $B_a$ is not parallel to $B_a$); it can destroy the geodesic
motion of the matter even in the absence of pressure.

%%%%%%%%%%%%%%%%%%%%%%%%%%%%%%%%
\subsection{Kinematic equations}
%%%%%%%%%%%%%%%%%%%%%%%%%%%%%%%%

Evolution of the expansion is governed
by the Raychaudhuri equation
\begin{equation}
\dot{\Theta}+{\textstyle{1\over3}}\Theta^2+
{\textstyle{1\over2}}(1+3w)\rho-\frac{c_{\rm a}^2}{3(1+w)}{\cal R}+
\frac{1}{2(1+w)a^2}\left(2c_{\rm s}^2\Delta+c_{\rm a}^2{\cal B}\right)-
\Lambda=0 \,,  \label{Ray}
\end{equation}
where $\Lambda$ is the cosmological constant,
${\cal R}=h^{ab}{\cal R}_{ab}$ is the projected curvature scalar, and
\[
\Delta=a\D^a\Delta_a\,,~~\Delta_a={a\D_a\rho\over\rho}\,,~~
{\cal B}={a^2\D^a{\cal B}_a\over B^2}\,,~~
{\cal B}_a=\D_aB^2\,,
\]
describe perturbations in the fluid and field energy densities.
Note that the overall magnetic effect includes a coupling to the projected
curvature, $c_{\rm a}^2{\cal R}$.

Magnetic influence on cosmic rotation is encoded in the
vorticity propagation equation
\begin{equation}
\dot{\omega}_a+2H\omega_a=-{\textstyle{1\over2}}
\mbox{curl}\,A_a \,, \label{v'}
\end{equation}
which may be rewritten, after eliminating the acceleration term via
Eq. (\ref{mdc}), as
\begin{equation}
\dot{\omega}_a+\left(2-3c_{\rm s}^2\right)H\omega_a=
-\frac{1}{2(1+w)\rho}B^b\D_b\mbox{curl}\,B_a \,.  \label{dotom}
\end{equation}
Thus there is a magnetically induced vorticity component
parallel to $\mbox{curl}\,B_a$. The effect
disappears if the directional derivative
$B^b\D_b\mbox{curl}\,B_a$ vanishes, i.e., when
$\mbox{curl}\,B_a$ does not change along the magnetic force
lines.

Kinematic anisotropies evolve via the shear propagation
equation
\begin{eqnarray}
\dot{\sigma}_{ab}+2H\sigma_{ab}&=&
-\frac{c_{\rm s}^2}{a(1+w)}\D_{\langle  a}\Delta_{b\rangle}-
\frac{1}{2(1+w)\rho}\D_{\langle  a}{\cal B}_{b\rangle}+
\frac{c_{\rm a}^2}{3(1+w)}{\cal R}_{\langle  ab\rangle}
\nonumber\\&\mbox{}&+
{\textstyle\frac{1}{2}}\pi_{ab}+
\frac{1}{(1+w)\rho}B^c\D_c\D_{\langle  a}B_{b\rangle}-
E_{ab} \,.  \label{dotsh}
\end{eqnarray}
The direct magnetic effects propagate through the field's
anisotropic stress ($\pi_{ab}$), as well as via anisotropies in the
distribution of magnetic energy
density ($\D_{\langle a}{\cal B}_{b\rangle}$)
 and of the field
vector itself ($\D_{\langle a}B_{b\rangle}$).
The latter effect vanishes when $\D_{\langle  a}B_{b\rangle}$
is invariant along the magnetic force lines. Also, the coupling
between the field and the projected curvature has led to an extra
magneto-geometrical contribution,
$c_{\rm a}^2{\cal R}_{\langle ab\rangle}$.

The kinematic quantities also obey constraint equations:
\begin{eqnarray}
\left(\mbox{div}\,\sigma\right)_a&=&
{\textstyle\frac{2}{3}}\D_a\Theta+\mbox{curl}\,\omega_a \,,  \label{shcon}\\
\mbox{div}\,\omega&=&0 \,,  \label{vorcon}\\
H_{ab}&=&\mbox{curl}\,\sigma_{ab}+
\D_{\langle  a}\omega_{b\rangle} \,.  \label{Hcon}
\end{eqnarray}

%%%%%%%%%%%%%%%%%%%%%%
\subsection{Curvature}
%%%%%%%%%%%%%%%%%%%%%%

The electric and magnetic Weyl tensors obey Maxwell-like
equations \cite{E2,mb}:
\begin{eqnarray}
\dot{E}_{ab}+3HE_{ab}-\mbox{curl}\,H_{ab} &=&
-{\textstyle{1\over2}}\rho(1+w)\sigma_{ab}+3H\pi_{ab}
\,,\label{gem1}\\
\dot{H}_{ab}+3HH_{ab}+\mbox{curl}\,E_{ab} &=&{\textstyle{1\over2}}
\mbox{curl}\,\pi_{ab}\,,\label{gem2}\\
\left({\rm div}\,E\right)_a&=&{\textstyle{1\over3}}\D_a\rho
+{\textstyle{1\over6}}{\cal B}_a
-{\textstyle{1\over2}}\left({\rm div}\,\pi\right)_a
\,,\label{gem3}\\
\left({\rm div}\,H\right)_a&=&(1+w)\rho\omega_a\,,\label{gem4}
\end{eqnarray}
where we have used Eq. (\ref{pi}). For a magnetized fluid the
projected curvature tensor ${\cal R}_{ab}$ is not in general
symmetric, but has the form ${\cal R}_{ab}={\cal R}_{\langle
ab\rangle}+{\textstyle{1\over3}}{\cal
R}h_{ab}+\varepsilon_{abc}{\cal R}^c$, where \cite{TB}
\begin{eqnarray}
{\cal R}_{\langle ab\rangle}&=&
\D_{\langle a}A_{b\rangle}-
\frac{1}{a^3}\left(a^3\sigma_{ab}\right)^{\rd}+\pi_{ab}\,,\label{3R_ab}\\
{\cal R}_a &=&\mbox{curl}\,A_a
-\frac{1}{a^3}\left(a^3\omega_{a}\right)^{\rd}\,,\\
{\cal R}&=&2\left(\rho-{\textstyle{1\over3}}\Theta^2
+\Lambda\right)\label{3R''}
\end{eqnarray}
Note that the background relation $3H^2=\rho+\Lambda$ ensures that
${\cal R}$ vanishes in the background, so that it is gauge-invariant.
By Eq. (\ref{v'}), the vector part of ${\cal R}_{ab}$ is simply
%\[%\begin{equation}
${\cal R}_a=-H\omega_a$,
%\]%\label{3R'}\end{equation}
which vanishes when the vorticity vanishes.
%The projected Ricci scalar evolves as \cite{TB}
%\begin{equation}
%\dot{{\cal R}}+2H{\cal R}=
%\frac{2H}{(1+w)a^2}\left[2c_{\rm s}^2\Delta+c_{\rm a}^2{\cal B}\right]
%\,.  \label{dotcR}\end{equation}
The dimensionless curvature perturbation ${\cal K}=a^2{\cal R}$
has comoving gradient \cite{TB}
\begin{equation}
a\D_a{\cal K}=2\rho a^2\Delta_a+a^3{\cal B}_a-4Ha^3\D_a\Theta\,,
\label{r}\end{equation}
which evolves as \cite{TB}
\begin{equation}
\left(a\D_a{\cal K}\right)^{\rd}=
{2aH\over\rho(1+w)}\D^2\left( 2\rho c_{\rm
s}^2\Delta_a+a{\cal B}_a\right)
+24a^3H^2c_{\rm s}^2{\rm curl}\,\omega_a\,.
\label{r'}\end{equation}

%%%%%%%%%%%%%%%%%%%%%%%%%%%%%%%%%%%%%%%%%
\subsection{Evolution of inhomogeneities}
%%%%%%%%%%%%%%%%%%%%%%%%%%%%%%%%%%%%%%%%%

The key gauge-invariant quantities describing inhomogeneity
are the comoving spatial gradients of the fluid
density, the field density,
the expansion and the field vector:
\begin{equation}
\Delta_a={a\D_a\rho\over\rho}\,,~~
{\cal B}_a=\D_aB^2\,,~~
\Theta_a=a\D_a\Theta\,,
~~B_{ab}=a\D_bB_a\,. \label{grad}
\end{equation}
Their propagation
equations are \cite{TB}
\begin{eqnarray}
\dot{\Delta}_a&=&3wH\Delta_a+(1+w)\Theta_a+
\frac{3aH}{\rho}\varepsilon_{abc}B^b\mbox{curl}\,B^c \,,
\label{dotDel_a}\\ \dot{\Theta}_a&=&-2H\Theta_a-
{\textstyle{1\over2}}{\rho}\Delta_a- \frac{c_{\rm
s}^2}{1+w}\D^2\Delta_a- {\textstyle{1\over2}}{a}{\cal B}_a-
\frac{a}{2(1+w)\rho}\D^2{\cal B}_a \nonumber\\&\mbox{}&+
{\textstyle\frac{3}{2}}a\varepsilon_{abc}B^b\mbox{curl}\,B^c
-\left[6c_{\rm s}^2+\frac{4c_{\rm
a}^2}{1+w}\right]aH\mbox{curl}\,\omega_a \,,\label{dotThe_a}\\
\dot{B}_{ab}&=&-2HB_{ab}+ \frac{c_{\rm
s}^2H}{1+w}\left[3B_{\langle a}\Delta_{b\rangle}+
B_{[a}\Delta_{b]}\right]-B_{\langle a}\Theta_{b\rangle}-
B_{[a}\Theta_{b]} \nonumber\\&&{}
+aB^c\D_c\sigma_{ab}+a\varepsilon_{ab}{}{}^dB^c\D_{\langle
c}\omega_{d\rangle}-aB_{[a}{\rm
curl}\,\omega_{b]}-a\varepsilon_{acd}B^cH^d{}_b \,.
\label{dotB_ab}
\end{eqnarray}
Note how the magnetic field couples to the magnetic Weyl
curvature via the last term in Eq. (\ref{dotB_ab}).

By eliminating the expansion gradients from the time derivative of
Eq. (\ref{dotDel_a}),
we arrive at \cite{TB}
\begin{eqnarray}
\ddot{\Delta}_a&=&-\left(2+3c_{\rm s}^2-6w\right)H\dot{\Delta}_a+
{\textstyle{1\over2}}\left[\left(1-6c_{\rm s}^2+8w-3w^2\right)\rho
-2\left(3c_{\rm s}^2-5w\right)\Lambda\right]\Delta_a
\nonumber\\&&{}+ c_{\rm s}^2\D^2\Delta_a +\frac{a}{2\rho}\D^2{\cal
B}_a+ {3a\over\rho}\left[\left(c_{\rm s}^2-w\right)\rho+
\left(1+c_{\rm
s}^2\right)\Lambda\right]\varepsilon_{abc}B^b\mbox{curl}\,B^c
\nonumber\\&&{}+ {\textstyle{1\over2}}(1+w)a{\cal B}_a+
\left(\frac{3aH}{\rho}\right)\varepsilon_{abc}B^b\mbox{curl}\,
\dot{B}^{\langle  c\rangle}+ \left[6(1+w)c_{\rm s}^2+4c_{\rm
a}^2\right]aH\mbox{curl}\,\omega_a \,.\label{ddotDel_a}
\end{eqnarray}
In the Newtonian limit, Eq. (\ref{ddotDel_a}) recovers the
results of \cite{RR}. The relativistic correction terms are the last three
terms on the right hand side, i.e., the terms with
${\cal B}_a$, $\varepsilon_{abc}B^b{\rm curl}\,\dot{B}^{\langle
c\rangle}$,
and ${\rm curl}\,\omega_a$.

%%%%%%%%%%%%%%%%%%%%%%%%%%%%%%%%%%%%%%%%%%
\section{Magnetized density perturbations}
%%%%%%%%%%%%%%%%%%%%%%%%%%%%%%%%%%%%%%%%%%

The equations (\ref{dotDel_a})--(\ref{dotB_ab})
provide the basis for a complete description of coupled
density-magnetic inhomogeneities.
We begin by isolating the evolution equations for the density
perturbation scalars $\Delta$
(a covariant alternative to the density contrast
 $\delta\rho/\rho$) and ${\cal B}$ (describing fluctuations in the magnetic
 energy density), and the curvature perturbation ${\cal K}$:
\[
\Delta=a\D^a\Delta_a\,,~~{\cal B}={a^2\D^a{\cal B}_a\over B^2}\,,~~
{\cal K}=a^2{\cal R}\,.
\]
The required evolution equations are \cite{TB}:
\begin{eqnarray}
\ddot{\Delta}&=&-\left(2+3c_{\rm
s}^2-6w\right)H\dot{\Delta}+{\textstyle{1\over2}}
\left[\left(1-6c_{\rm s}^2+8w-3w^2\right)\rho-2 \left(3c_{\rm
s}^2-5w\right)\Lambda\right]\Delta \nonumber\\&&{} +c_{\rm
s}^2\D^2\Delta -{\textstyle{1\over2}} \left[\left(1-3c_{\rm
s}^2+2w\right)\rho-\left(1+3c_{\rm s}^2
\right)\Lambda\right]c_{\rm a}^2{\cal B}+{\textstyle{1\over2}}
c_{\rm a}^2\D^2{\cal B} \nonumber\\&&{}+{\textstyle{1\over3}}
\left[\left(2-3c_{\rm s}^2+3w\right)\rho-\left(1+3c_{\rm
s}^2\right) \Lambda\right]c_{\rm a}^2{\cal K} \,,\label{ddotDel}\\
\dot{{\cal B}}&=& \frac{4}{3(1+w)}\dot{\Delta}+ \frac{4(c_{\rm
s}^2-w)H}{1+w}\Delta \,, \label{dotcB}\\ \dot{{\cal K}}&=&
\frac{4c_{\rm s}^2H}{1+w}\Delta +\frac{2c_{\rm a}^2H}{1+w}{\cal B}
\,.  \label{dotcK}
\end{eqnarray}
This system of equations governs the coupling
between density fluctuations $\Delta$,
magnetic fluctuations ${\cal B}$ and curvature fluctuations
${\cal K}$. Eq. (\ref{dotcK}) shows that ${\cal K}$ grows
if $\Delta$ and ${\cal B}$ are growing, while Eq. (\ref{dotcB})
shows that if $c_{\rm s}^2\geq w$ (which holds in the radiation
and dust eras), then ${\cal B}$ grows in concert with growing
$\Delta$.

The magnetic field introduces a direct effect, via the term
$c_{\rm a}^2{\cal K}$ in Eq. (\ref{ddotDel}),
of the curvature on the density perturbations.
In the non-magnetized case, there are two modes of $\Delta$, which
is governed by the single second-order equation (\ref{ddotDel})
with $c_{\rm a}=0$. In this case, the evolution of $\Delta$ is
independent of ${\cal K}$, and ${\cal K}$ is determined once $\Delta$
is known, via Eq. (\ref{dotcK}).
Magnetism introduces
two additional modes, since the system has four degrees of
freedom.
These modes are nonadiabatic, and can source density perturbations,
i.e., even when $\Delta(t_0)=0=\dot{\Delta}(t_0)$, magnetic effects
will lead to $\Delta\neq0$ for $t>t_0$.
If one omits the magneto-curvature effect, then the evolution
equation for ${\cal K}$, Eq. (\ref{dotcK}), is uncoupled from
the system, which can then be decoupled via
a third-order equation in $\Delta$. Neglecting the magneto-curvature
effect thus removes one of the additional nonadiabatic modes.

For zero
cosmological constant, we can solve the system analytically
 in the radiation and dust eras,
treating super- and sub-horizon scales separately.
Some solutions
were given in \cite{TB}. There, however,  magneto-curvature
effects were neglected in three out of the four cases. Here, we
generalize some of the solutions to
incorporate the magneto-curvature coupling, and we show that the
magneto-curvature coupling cannot in general be neglected, since
it leads to important qualitative differences in the behavior of
$\Delta$.

%%%%%%%%%%%%%%%%%%%%%%%%%%
\subsection{Radiation era}
%%%%%%%%%%%%%%%%%%%%%%%%%%

During the radiation era, $w=c_{\rm s}^2={1\over3}$,
$\rho=\rho_0(a_0/a)^4$,
and the Alfv\'{e}n speed does not change along
the fluid flow, i.e., $\dot{c}_{\rm a}=0$, reflecting the radiation-like
evolution of the magnetic energy density, as given by Eq. (\ref{B^2ev}).
For the Fourier modes
with comoving wave-number $k$, we get
\begin{eqnarray}
\left({a\over a_0}\right)^2\Delta'' &=&
\left[2-{\textstyle{1\over3}}\left(\frac{ k}{k_{{\rm h}0}}\right)^2
\left({a\over a_0}\right)^2\right]
\Delta-
\left[1+{\textstyle{1\over2}}
\left(\frac{ k}{k_{{\rm h}0}}\right)^2\left({a\over a_0}\right)^2
\right]c_{\rm a}^2{\cal B}+
2c_{\rm a}^2{\cal K} \,,  \label{rnddotDel}\\
{\cal B}'&=& \Delta'\,,\label{rn'}\\
\left({a\over a_0}\right){\cal K}'&=&\Delta+
{\textstyle{3\over2}}c_{\rm a}^2{\cal B} \,,  \label{rndotcB-dotcK}
\end{eqnarray}
where a prime denotes $d/d(a/a_0)$ and $k_{{\rm h}0}=a_0H_0$ is
the comoving wavenumber of the horizon at $a_0$.

\subsubsection{Superhorizon scales and the curvature coupling}

In the long wavelength limit $ k \ll k_{{\rm h}0}$, the
system has the power-law solution
\begin{eqnarray}
\Delta&=&C_{(0)}+
\sum_{\alpha}C_{(\alpha)}\left({a\over a_0}\right)^\alpha
\,,\label{lrnDel}\\
{\cal B} &=& -\left({2\over3 c_{\rm a}^2}\right)C_{(0)}+
\sum_{\alpha}C_{(\alpha)}\left({a\over a_0}\right)^\alpha
\,,\label{lrn'}\\
{\cal K} &=&-\left({4\over3 c_{\rm
a}^2}\right)C_{(0)}+\left(1+{\textstyle{3\over2}}c_{\rm
a}^2\right)
\sum_{\alpha}{C_{(\alpha)}\over\alpha}\left({a\over a_0}\right)^\alpha
 \,, \label{lrn''}
\end{eqnarray}
where $C_{(0)}$
and $C_{(\alpha)}$ are constants, and
the parameter $\alpha$ satisfies the cubic equation
\begin{equation}
\alpha^3-\alpha^2-\left(2-c_{\rm a}^2\right)\alpha
-(2+3c_{\rm a}^2)c_{\rm a}^2=0 \,.
\label{lrDelz}
\end{equation}
The cubic has one positive and two negative roots.
One of the negative roots corresponds to a decaying
nonadiabatic mode. The other
nonadiabatic mode is the $C_{(0)}$-mode, which is constant.
The remaining cubic roots correspond to the magnetized versions of
the standard adiabatic modes, one growing and one decaying.
Since $c_{\rm a}^2$ is small, we can find the roots perturbatively.
The zero-order roots are $0,-1,2$ (the $\alpha=0$ solution is
spurious in the non-magnetized case). To lowest order, we find
that:
\begin{equation}
\alpha=\left\{\begin{array}{r}
0-c_{\rm a}^2+O\left(c_{\rm a}^4\right)\,,\\
\\
-1+c_{\rm a}^2+O\left(c_{\rm a}^4\right)\,,\\
\\
2+{\textstyle{1\over2}}c_{\rm a}^4
+O\left(c_{\rm a}^6\right)\,.
\end{array}\right.
\label{nc}\end{equation}
Thus the adiabatic growing mode of the non-magnetized case is
slightly {\em enhanced} by magnetic effects
(the enhancement is not felt to lowest order in $c_{\rm a}^2$);
the adiabatic decaying mode
decays less rapidly by virtue of magnetic effects; the
decaying nonadiabatic mode decays very slowly;
and the final, nonadiabatic, mode is constant. To lowest order
\begin{equation}
\Delta= C_{(+)}\left({a\over a_0}\right)^2+C_{(1-)}\left({a\over
a_0}\right)^{-1+c_{\rm a}^2}+C_{(0)}+
C_{(2-)}\left({a\over
a_0}\right)^{-c_{\rm a}^2}
\,.\label{new}
\end{equation}
The magnetic and curvature fluctuations are
given by equations (\ref{lrn'}) and (\ref{lrn''}), with $\alpha$
given by Eq. (\ref{nc}).

This new solution in Eq. (\ref{new}) can be compared with the
solution that arises when the magneto-curvature coupling
term $c_{\rm a}^2{\cal K}$ is
ignored in Eq. (\ref{lrnDel}) \cite{TB}. Then the last term in Eq.
(\ref{lrDelz}) falls away, leading to the quadratic
$\alpha^2-\alpha-\left(2-c_{\rm a}^2\right)=0$. To lowest order
\[
\alpha=\left\{\begin{array}{r}
-1+{\textstyle{1\over3}}c_{\rm a}^2 \,,\\
\\
2-{\textstyle{1\over3}}c_{\rm a}^2 \,,
\end{array}\right.
\]
so that the density perturbation is given by
\[%\begin{equation}
\Delta=C_{(+)}\left({a\over a_0}\right)^{2-{1\over3}c_{\rm a}^2}+
C_{(-)}\left({a\over a_0}\right)^{-1+{1\over3}c_{\rm a}^2}
+C_{(0)}\,,
\]%  \label{lrnDel1}\end{equation}
and the magnetic fluctuations are
\[
{\cal B} = \Delta -\left[1-\left({2\over c_{\rm
a}^2}\right)\right]C_{(0)}\,.
\]

Clearly, omitting the magneto-curvature coupling has a significant
qualitative impact. Not only is one of the nonadiabatic modes
($C_{(2-)}$) removed, as expected, but we also find that the
growing mode is slightly {\em damped}, at odds with the correct
solution in Eq. (\ref{new}). Thus the magneto-curvature coupling,
which was identified in general in \cite{TB}, turns out to have a
crucial role in increasing (even though it is only by a small
amount) the standard adiabatic modes of density perturbations on
large scales in the radiation era. It is not reasonable to omit
the magneto-curvature coupling in this case.

\subsubsection{Subhorizon scales and magneto-sonic waves}

At the opposite end of the wavelength spectrum, when
$k \gg k_{{\rm h}0}$, we differentiate
Eq. (\ref{rnddotDel}) and use
Eq. (\ref{rndotcB-dotcK}) to decouple the system. Integrating once
we get
\[%\begin{equation}
6\left({a\over a_0}\right)^2\Delta''+
2\left(\frac{ k}{k_{{\rm h}0}}\right)^2\left({a\over a_0}\right)^2
\left(1+{\textstyle{3\over2}}
c_{\rm a}^2\right)\Delta= 6C_{{\cal K}}-
3C_{{\cal B}}c_{\rm a}^2\left({k\over k_{{\rm
h}0}}\right)^2
\left({a\over a_0}\right)^2 \,,
\]% \label{srnddotDel1}\end{equation}
where $C_{{\cal K}}$ is an additional constant associated with
curvature effects. (We have ignored higher
order terms in $c_{\rm a}^2$, given the
weakness of the magnetic field.) This has
solution (to lowest order in $c_{\rm a}^2$)
\begin{eqnarray}
\Delta&=&\left[C_{(1)}-C_{{\cal K}}
{\rm Si}\left(\beta {k\over k_{{\rm h}0}}{a\over a_0}\right)
\right]\sin\left(\beta {k\over k_{{\rm h}0}}{a\over a_0}\right)\nonumber\\
&&{}+\left[C_{(2)}-C_{{\cal K}}
{\rm Ci}\left(\beta {k\over k_{{\rm h}0}}{a\over a_0}\right)
\right]\cos\left(\beta {k\over k_{{\rm h}0}}{a\over a_0}\right)
-C_{{\cal B}}c_{\rm a}^2
\,,  \label{srnDel}
\end{eqnarray}
where $C_{(i)}$ are constants, Si and Ci are the sine and cosine
integral functions,\footnote{
${\rm Si}(x)=\int_0^xt^{-1}\sin t\,dt$ and
${\rm Ci}(x)=\gamma+\ln x+\int_0^xt^{-1}(\cos t-1)dt$,
where $\gamma= 0.578\cdots$ is Euler's constant  \cite{AS}.
}
and
\begin{equation}
\beta=c_{\rm s}\left(1+{\textstyle{3\over4}}c_{\rm a}^2\right)
\,,  \label{srDelz}
\end{equation}
where $c_{\rm s}=1/\sqrt{3}$ is the adiabatic sound speed. Thus
$\beta$ is the magnetized (nonadiabatic) sound speed of
magneto-sonic waves. These waves differ slightly in amplitude and
frequency from the adiabatic acoustic waves.

The  magneto-curvature
coupling, reflected in the nonadiabatic $C_{{\cal K}}$ mode, has the effect of
slightly modulating the amplitude of acoustic oscillations, with
the effect decreasing as $a/a_0$ increases.
The main magnetic effect is on the frequency.
Comparing our result in Eq. (\ref{srDelz})
to the standard solutions of
magnetic-free models (see, e.g., \cite{P}),
 we see that the field has
increased the frequency of acoustic oscillations.
Since $a\propto\sqrt{t}$, the magnetized acoustic frequency is
\begin{equation}
\nu_{\rm ac,mag}=\nu_{\rm ac}\left(1+{\textstyle{3\over2}}c_{\rm a}^2
\right)~\mbox{ where }~\nu_{\rm ac}={H_0\over3\pi }\left({k\over
k_{{\rm h}0}}\right)^2\,.
\label{3}\end{equation}
This magnetic correction results from the
the ``tensioning" effect of magnetic force lines in the plasma,
which produces a nonadiabatic increase of the sound speed via a
contribution from the Alfv\'{e}n speed.
As a result, the magnetic influence brings
the acoustic peaks of short-wavelength radiation density
oscillations closer, producing in principle an
observable signature on CMB anisotropies \cite{adgr}. An additional
effect comes from the nonadiabatic constant mode in Eq. (\ref{srnDel}). Its
presence suggests that the average value of the density contrast
is {\em nonzero}, unlike the magnetic-free case.

%%%%%%%%%%%%%%%%%%%%%
\subsection{Dust era}
%%%%%%%%%%%%%%%%%%%%%

After recombination, in a baryon-dominated cold matter
background, $w=0=c_{\rm s}^2$, $a=a_0 (t/t_0)^{2/3}$,
$H=2/3t$ and $\rho=4/3t^2$. The Alfv\'{e}n speed is no
longer constant, but by Eq. (\ref{B^2ev})
varies as
\[%\begin{equation}
c_{\rm a}^2=\left(c_{\rm a}^2\right)_0 \left({t_0\over t}\right)^{2/3}\,,
\]%\label{cd}\end{equation}
reflecting the fact that the magnetic
energy density drops faster than that of nonrelativistic
matter. Thus magnetic effects grow weaker as the expansion of the
universe proceeds beyond recombination.

The equations for the Fourier modes become
\begin{eqnarray}
\Delta''&=&-{\textstyle{4\over3}}\left({t_0\over t}\right)\Delta'+
{\textstyle{2\over3}}\left({t_0\over t}\right)^2\Delta\nonumber\\
&&{}-
{\textstyle{2\over3}}\left(c_{\rm a}^2\right)_0\left({t_0\over t}\right)^{8/3}\left[
1+{\textstyle{1\over3}}
\left(\frac{k}{k_{{\rm h}0}}\right)^2\left({t\over t_0}\right)^{2/3}
\right]{\cal B}+
{\textstyle{8\over9}}
\left(c_{\rm a}^2\right)_0\left({t_0\over t}\right)^{8/3}{\cal K}
\,,  \label{dnddotDel} \\
{\cal B}' &=& {\textstyle{4\over3}}\Delta' \,, \label{dn'} \\
{\cal K}' &=& {\textstyle{4\over3}}
\left(c_{\rm a}^2\right)_0\left({t_0\over t}\right)^{5/3}
{\cal B}\,,  \label{dndotcB-dotcK}
\end{eqnarray}
where a prime denotes $d/d(t/t_0)$.
Thus
\begin{eqnarray}
{\cal B}&=&{\textstyle{4\over3}}\left(\Delta+C_{{\cal B}}\right)\,,
\label{dnc'} \\
{\cal K}'&=&{\textstyle{16\over9}}
\left(c_{\rm a}^2\right)_0\left({t_0\over t}\right)^{5/3}
\left[\Delta+C_{{\cal B}}\right] \,,  \label{dncB-dotcK}
\end{eqnarray}
where $\dot{C}_{{\cal B}}=0$. We can now decouple the system.

\subsubsection{Superhorizon scales}

For long wavelength fluctuations, we get
\begin{equation}
9\left({t\over t_0}\right)^3\Delta'''+36\left({t\over t_0}\right)^2
\Delta''+14\left({t\over t_0}\right)\Delta'-
4\Delta =0 \,,  \label{ldndddotDel}
\end{equation}
to lowest order in $c_{\rm a}^2$. Note that curvature effects are
quadratic in $c_{\rm a}^2$ and do not contribute at this level. In
fact equations (\ref{dnddotDel}), (\ref{dndotcB-dotcK}) guarantee
that, to lowest order in $c_{\rm a}^2$, curvature has no effect on
magnetised disturbances in the dust distribution. We can solve Eq.
(\ref{ldndddotDel}), which is of Euler-type:
\begin{equation}
\Delta=C_{(+)}\left({t\over t_0}\right)^{2/3}+C_{(1-)}
\left({t\over t_0}\right)^{-1}+
C_{(2-)}\left({t\over t_0}\right)^{-2/3} \,.  \label{lldnDel}
\end{equation}
Thus, the field has simply added the nonadiabatic decaying mode $C_{(2-)}$
to the evolution of superhorizon
density perturbations, while the non-magnetized adiabatic modes
are unchanged. The growth of large-scale
matter aggregations proceeds virtually unaffected by the presence
of the field or by curvature complexities.

Magnetic effects on superhorizon scales
in the dust era do not change the adiabatic growing mode to lowest order in
$c_{\rm a}^2$.
The adiabatic decaying mode is also unchanged (unlike the radiation
case).
However, a new nonadiabatic decaying mode arises, which decays less
rapidly than the adiabatic mode.

\subsubsection{Subhorizon scales}

On subhorizon scales
\begin{eqnarray*}
9\left({t\over t_0}\right)^3\Delta'''+36\left({t\over t_0}\right)^2
\Delta''+
14\left({t\over t_0}\right)\left[1+{\textstyle{4\over21}}
\left(c_{\rm a}^2\right)_0\left({k\over k_{{\rm h}0}}\right)^2
\right]\Delta'&& \nonumber\\
{}-4\left[1-{\textstyle{4\over9}}\left(c_{\rm a}^2\right)_0
\left({k\over k_{{\rm h}0}}\right)^2\right]
\Delta =-{\textstyle{16\over9}}\left(c_{\rm a}^2\right)_0
\left({k\over k_{{\rm h}0}}\right)^2C_{{\cal B}} \,,~&&
%\label{sdndddotDel}
\end{eqnarray*}
where again we have ignored terms of higher order in $(c_{\rm a}^2)_0$. The
solution is
\begin{eqnarray}
\Delta&=&C_{(+)}\left({t\over t_0}\right)^{\alpha_+}+
C_{(-)}\left({t\over t_0}\right)^{\alpha_-}+
C_{({\cal B}-)}\left({t\over t_0}\right)^{-2/3}\nonumber\\
&&{}+
C_{\cal B}\left[\frac{4\left(c_{\rm a}^2\right)_0 k ^2}
{4\left(c_{\rm a}^2\right)_0 k ^2-9k_{{\rm h}0}^2}
\right] \,,  \label{sdnDel}
\end{eqnarray}
with
\begin{equation}
\alpha_\pm={\textstyle{1\over6}}\left[-1\pm5\sqrt{1-{\textstyle{32\over75}}
\left(c_{\rm a}^2\right)_0\left({k \over k_{{\rm h}0}}\right)^2 }\,\right] \,.
\label{sdnz12}
\end{equation}
The magnetic influence is expressed in two ways: additional decaying
($C_{({\cal B}-)}$) and constant ($C_{{\cal B}}$) nonadiabatic modes; and
modification of the non-magnetized adiabatic modes ($C_{(\pm)}$).
The net effect is to
inhibit the growth of matter aggregations, as noted also in
the Newtonian case \cite{RR}. Note that the magnetic effects, direct
or indirect, become less important after matter-radiation equality,
due to the decrease of the Alfv\'{e}n speed.
The damping of the growing mode is greater on smaller scales.
Indeed there is a minimum scale, below which the solution in Eq.
(\ref{sdnDel}) oscillates, since the magnetic pressure balances
gravitational infall. This magnetic Jeans scale follows from Eq.
(\ref{sdnz12}):
\begin{equation}
\lambda_{\rm mJ}(t_0)={\textstyle{4\over5}}\pi\sqrt{6}\lambda_{\rm
a}(t_0)\,,\label{mj}
\end{equation}
where
\[%\begin{equation}
\lambda_{\rm a}=c_{\rm a}t={\textstyle{2\over3}}c_{\rm a}\lambda_{\rm h}
\]%\label{ah}\end{equation}
is the Alfv\'{e}n horizon, with $\lambda_{\rm h}=H^{-1}$ the Hubble
scale.
In fact, given the weakness of the magnetic field, it is likely
that kinetic pressure cannot be ignored near the magnetic Jeans
scale. In this case, a more sophisticated analysis is necessary,
to incorporate nonrelativistic pressure effects in baryonic matter.
On scales well above the Alfv\'{e}n horizon (equivalently,
magnetic Jeans scale) but well within the Hubble horizon, i.e.,
for
\[
k_{{\rm h}0}\ll k\ll k_{{\rm a}0}~\mbox{ where }~k_{\rm
a}={3\over2c_{\rm a}}k_{\rm h}\,,
\]
we find that the magnetized corrections $\alpha_\pm$ of the
adiabatic exponents are
\[
\alpha_\pm=\left\{\begin{array}{r}
{\textstyle{2\over3}}-{\textstyle{2\over5}}(k/ k_{{\rm a}0})^2\,,\\
\\
-1+{\textstyle{2\over5}}(k/ k_{{\rm a}0})^2\,.\\
\end{array}\right.
\]

The way in which magnetic effects act to increase the adiabatic
Jeans length may be qualitatively understood as follows.
Consider a tube of magnetic
force-lines with instantaneous cross-sectional area
$\delta S$.
In a perfectly conducting medium the field remains
frozen into the fluid, i.e., the magnetic force-lines
always connect the same particles \cite{E2}.
More precisely, the induction equation (\ref{Max2}) shows that
$a^3B^a$ is a connecting vector. Thus the volume of the tube
is given by
\[
\delta V=\delta\ell\,\delta S\propto a^3B\,\delta S\,.
\]
However, we also have that in general, $\delta V\propto a^3$. It
follows that
\[
\left(B\,\delta S \right)^{\rd}=0\,.
\]
 The conservation law in Eq. (\ref{B^2ev}) shows that $B\propto
 a^{-1}$; thus
\[
\delta S\propto a^2\,,
\]
so that the cross section of the
flux tube increases as the expansion redshifts the energy density
of the field. Thus the field acts against gravitational infall.

%%%%%%%%%%%%%%%%%%%%%%%%%%%%%%%%%%%%%%%%%%%%%%%%%%%%%%%%%%%%%%%%%%%%
\subsection{Pure-magnetic and magnetized isocurvature perturbations}
%%%%%%%%%%%%%%%%%%%%%%%%%%%%%%%%%%%%%%%%%%%%%%%%%%%%%%%%%%%%%%%%%%%%

We have seen that the magnetic field introduces nonadiabatic modes in the
density perturbations.
This means that the field itself can generate fluctuations in the
density, even when there are no primordial density fluctuations. Thus, if
\begin{equation}
\Delta(t_0)=0=\dot{\Delta}(t_0)\,,
\label{pm}\end{equation}
where $t_0$ is the epoch of magnetogenesis in the early radiation era,
then nonzero $\Delta$
will arise purely from the magnetic field; in the absence of magnetogenesis,
Eq. (\ref{pm}) would imply $\Delta=0$ for $t>t_0$. These nonadiabatic pure
magnetic density perturbations can be found explicitly from the
solutions given above. On superhorizon scales
(assuming that the field is created on these scales at $t_0$), the general
solution in Eq. (\ref{new}) implies with the initial conditions in
Eq. (\ref{pm}) that
the pure-magnetic nonadiabatic mode is (to lowest
order in $c_{\rm a}^2$)
\begin{eqnarray}
\Delta_{\rm pm}& =&
-{\textstyle{1\over3}}\left[C_{(0)}+C_{(2-)}\right]
\left({a\over a_0}\right)^{2}\left[1+2
\left({a\over a_0}\right)^{-3+c_{\rm a}^2}\right]
\nonumber\\
&&{}+C_{(0)}+C_{(2-)}\left({a\over a_0}\right)^{-c_{\rm a}^2}\,.
\label{new'}
\end{eqnarray}
The pure-magnetic density perturbations have a
dominant growing mode of the same strength as
in the non-magnetized adiabatic case. The decaying modes are in
fact the isocurvature part of the pure-magnetic density
perturbations, as we now show.

Equation (\ref{pm}) is often taken to characterize isocurvature
perturbations, but it does so only in specific cases
\cite{ent}. For
magnetized perturbations, this is not the isocurvature condition.
Isocurvature density perturbations are those for which the
curvature perturbation of the initial hypersurface orthogonal to the fluid flow
is spatially constant, i.e.,
$(a\D_a{\cal K})(t_0)=0$. (There is also the implicit condition
that $\omega_a=0$, which is necessary for the existence of
the spatial hypersurface.) Taking the
the comoving divergence of Eqs. (\ref{r}) and (\ref{dotDel_a}),
we find the condition for magnetized isocurvature perturbations:
\begin{equation}
\dot{\Delta}+{\textstyle{3\over2}}(1-w)H\Delta={\textstyle{3\over4}}(1-w)
c_{\rm a}^2H{\cal B}-c_{\rm a}^2H{\cal K}~\mbox{ at }~t=t_0\,.\label{iso1}
\end{equation}
In the non-magnetized case $c_{\rm a}=0$, it is clear that this
condition is satisfied by Eq. (\ref{pm}), but when $c_{\rm a}>0$, then
Eq. (\ref{pm}) does not characterize isocurvature perturbations.

As an example, consider the implication of the magnetized
isocurvature condition in the dust era, on scales well above the
Alfv\'{e}n horizon but well within the Hubble horizon, i.e.,
$k_{{\rm h}0}\ll k\ll k_{{\rm a}0}$. Then Eqs.
(\ref{iso1}) and (\ref{sdnDel}) give, to lowest order,
\[
C_{(+)}={\textstyle{2\over5}}\left(c_{\rm a}^2\right)_0\left[C_{(-)}
+C_{{\cal B}}\right]-{\textstyle{1\over5}}C_{{\cal K}}\,.
\]

On superhorizon scales,
Eq. (\ref{iso1}) holds for all $t$ by virtue of Eq. (\ref{r'}),
which implies $(a\D_a{\cal K})^{\rd}=0$. The magnetized
isocurvature condition then selects a sub-class of the general
superhorizon solutions found above.
In the radiation era, we find
(to lowest order in $c_{\rm a}^2$)
\begin{equation}
\Delta_{\rm iso}=C_{(1-)}\left(\frac{t}{t_0}\right)^{-\frac{1}{2}c_a^2}+
C_{(2-)}\left(\frac{t}{t_0}\right)^{-\frac{1}{2}+\frac{1}{2}c_a^2} \,.
\label{iso4}\end{equation}
Equation (\ref{iso4}) arises from the general superhorizon
solution Eq. (\ref{new}) by eliminating the non-decaying modes.
In the dust era,
\begin{equation}
\Delta_{\rm iso}=C_{(1-)}\left(\frac{t}{t_0}\right)^{-1}+
C_{(2-)}\left(\frac{t}{t_0}\right)^{-\frac{2}{3}} \,,
\label{iso5}\end{equation}
which is a special case of the general superhorizon dust solution
Eq. (\ref{lldnDel}), once again without the growing mode.

It is clear from Eqs. (\ref{iso4}) and (\ref{iso5}) that
magnetized isocurvature perturbations on superhorizon scales
are purely decaying. They are
very different from the pure-magnetic solution Eq. (\ref{new'}) that is based on
the initial conditions in Eq. (\ref{pm}). The latter has a
constant and a growing mode. Magnetized nonadiabatic perturbations
on superhorizon scales can contribute to the growing mode and
generate a constant mode, whereas the
magnetized isocurvature perturbations are purely decaying.

%%%%%%%%%%%%%%%%%%%%%%%%%%%%%%%%%%%%%%%%%%%%%%%%%%%%%%%%%
\section{Magnetized cosmic vortices and Alfv\'{e}n waves}
%%%%%%%%%%%%%%%%%%%%%%%%%%%%%%%%%%%%%%%%%%%%%%%%%%%%%%%%%

In the previous section, we
generalized the results given in \cite{TB}, which itself
provided a relativistic extension of previous work on
magnetized density perturbations.
A general inhomogeneous perturbation
is characterized not only by its magnitude, i.e. the density
perturbation $\Delta$, but also by its rotation and
deformation properties, as described in general terms in \cite{EB2,MT}.
Recently, these properties were investigated in CDM,
and it
was shown how the small stresses (isotropic and anisotropic)
from residual velocity dispersion can have an important effect
on rotation and deformation, even though the effect on density
perturbations is effectively negligible \cite{mtm}.

The evolution equations for rotational
and deformation variables in an imperfect fluid were derived in \cite{MT}.
The evolution equations for inhomogeneities were coupled
to causal transport equations for viscosity and heat conduction.
By Eq. (\ref{emt'}), a magnetized perfect
fluid can be considered as an imperfect fluid with
anisotropic stress, and the equations of \cite{MT}
may be specialized to this case. However, the system needs to
be completed by evolution equations for the magnetic
stress, which are
determined by Maxwell's equations. Here we investigate the coupled
equations governing rotational and deformational inhomogeneity
in the fluid and magnetic field.

The comoving gradient of the density inhomogeneity $\Delta_a$
splits irreducibly as
\[%\begin{equation}
a\D_b\Delta_a=
{\textstyle{1\over3}}\Delta h_{ab}+
\varepsilon_{abc}W^c+
\xi_{ab} \,.
\]%\label{Del_ab}\end{equation}
The density perturbation is the comoving divergence,
the rotational part is given by the comoving curl and the
deformation part is the comoving PSTF derivative:
\[%\begin{equation}
\Delta=a\D^a\Delta_a\,,~~
W_a=-{\textstyle{1\over2}}a\,{\rm curl}\,\Delta_{a}\,,~~
\xi_{ab}=a\D_{\langle  a}\Delta_{b\rangle}\,.
\]% \label{W_a-xi_ab}\end{equation}
The vector $W_a$ governs
rotational instabilities in the density distribution of the
matter, and ${\rm div}\, W=0$.
On the other hand, $\xi_{ab}$ determines the volume-true
anisotropic distortion, with $h^{ab}\xi_{ab}=0$.
Both quantities describe differential, i.e., infinitesimal,
properties. Here we focus on rotation, and in the next section
we look at anisotropic deformations.

A fundamental property of rotational perturbations is that they
are proportional to the vorticity vector. This arises from
the identity Eq. (\ref{id2})
which ensures that the curl of any gradient field, such as $\Delta_a$,
derives from the vorticity. It follows that
\begin{equation}
W_a=-3a^2H(1+w)\omega_a\,.
\label{w}\end{equation}
Not only $W_a$,
but also rotational perturbations in
magnetic density and expansion inhomogeneities, are parallel to the vorticity:
\[
{\rm curl}\,{\cal B}_a\equiv{\rm curl}\,\D_aB^2={4\over3(1+w)}W_a\,,~
{\rm curl}\,\Theta_a\equiv{\rm curl}\,a\D_a\Theta=-{\dot{H}\over(1+w)H}W_a\,.
\]
Thus all rotational instability in the magnetized medium arises from
vorticity. As we have seen from the vorticity propagation
equation (\ref{dotom}), magnetic inhomogeneities can source vorticity.
Rewriting Eq. (\ref{dotom}) using Eq. (\ref{w}), we have
\begin{equation}
\dot{W}_a+{\textstyle{3\over2}}(1-w)HW_a=\left(
\frac{3a^2H}{2\rho}\right)B^b\D_b\mbox{curl}\,B_a \,.  \label{dotW_a}
\end{equation}
Thus the field is a source of vorticity provided that its curl varies
along its force lines.
Furthermore, the effect of the field is to induce precession of
the rotational vector $W_a$. In the absence of the field, $\dot{W}_a$
remains parallel to $W_a$, so that the initial direction is
preserved along the fluid flow. By contrast, in the magnetized
case, $\dot{W}_a$ is no longer parallel to $W_a$, and the initial
direction changes along the fluid flow. The rate of precession is
\begin{equation}
\nu_{\rm prec}={|B^b\D_b\mbox{curl}\,B_a|\over2(1+w)\rho|\omega_a|}\,.
\label{prec}
\end{equation}

Equation (\ref{dotW_a}) shows how the magnetic field can become a
source of density vortices. However, the field effect upon
pre-existing rotational perturbations is not clear yet. To
quantify the magnetic influence on $W_a$ we need to go one step
further and obtain a decoupled equation for the
evolution of $W_a$.
We take the curl of Eq. (\ref{ddotDel_a}), using the above results and the
identities in Eqs. (\ref{id2}), (\ref{id1}) and (\ref{id3}),
and we arrive at the required evolution equation (with $\Lambda=0$):
\begin{equation}
\ddot{W}_a+\left(4-3w\right)H\dot{W}_a+
{\textstyle{1\over2}}\left[1-7w+3c_{\rm s}^2(1+w)\right]\rho W_a=
\left[\frac{c_{\rm a}^2}{3(1+w)}\right]\D^2W_a \,.  \label{ddotW_a}
\end{equation}
This is a wave equation for $W_a$, with signal speed $v_{\rm a}$ given
by\footnote{
A similar equation was derived in \cite{MT}
for a fluid with shear viscosity; in that case
$v^2=\eta/[\tau\rho(1+w)]$, where $\eta$ is the viscosity and $\tau$
is the causal relaxation time.
}
\[
  v_{\rm a}^2={c_{\rm a}^2\over 3(1+w)}\,.
\]
Propagating solutions of this equation are Alfv\'{e}n waves
(compare \cite{cmb}), i.e., incompressible, vector waves, as
opposed to the compressible, scalar magneto-sonic waves.
In the non-magnetized case, the signal speed vanishes, and no wave
solutions exist.
We note also that the only scale-dependence arising in the wave equation is
via the magnetic $c_{\rm a}^2$ term. Thus magnetized vortices are
scale-dependent, unlike the non-magnetized case.

Decomposing the solenoidal vector $W_a$ into Fourier modes
${\cal W}$, Eq. (\ref{ddotW_a}) gives
\begin{equation}
\ddot{\cal W}+\left(4-3w\right)H\dot{\cal W}+
\left\{{\textstyle{1\over2}}
\left[1-7w+3c_{\rm s}^2(1+w)\right]\rho+
\frac{ c_{\rm a}^2k^2}{3(1+w)a^2}\right\}{\cal W}=0\,.  \label{nddotW_a}
\end{equation}
Clearly, $W_a$ can only grow if the term in square brackets becomes
negative. If $\dot{w}=0$, as in the radiation and dust eras, then the quantity
$1-7w+3(1+w)c_{\rm s}^2=(3w-1)(w-1)$
becomes negative when ${1\over3}<w<1$.
Thus it is only when matter stiffer than
radiation dominates the universe (and is coupled to the magnetic field),
that vortices in the density distribution can grow (in the linear
regime).
Furthermore, the presence of the
magnetic field ensures that such growth occurs only on scales larger
than the critical ``rotational Jeans" wavelength
\[%\begin{equation}
\lambda_{\rm rJ}=2\pi c_{\rm a}\left[{\frac{2}{3(3w-1)(1-w^2)\rho}}
\right]^{1/2} \,,
\]%\label{lambda_c}\end{equation}
which is small due to the weakness of the field.

In the radiation era, Eq. (\ref{nddotW_a})
becomes
\[%\begin{equation}
{\cal W}''+{2}\left({a_0\over a}\right){\cal W}'+
\left({k\over k_{{\rm a}0}}\right)^2{\cal W}=0 \,,
\]%\label{rnddotW_a}\end{equation}
where a prime denotes $d/d(a/a_0)$, and $k_{\rm a}=2k_{\rm h}/c_{\rm a}$
is the wavenumber of the Alfv\'{e}n horizon
$\lambda_{\rm a}={1\over2}c_{\rm a}\lambda_{\rm h}$. This has
the general solution
\begin{equation}
{\cal W}=\frac{a_0}{a}
\left[C_{(1)}\cos\left(
{k\over k_{{\rm a}0}}\frac{a}{a_0}\right)
+C_{(2)}\sin\left(
{k\over k_{{\rm a}0}}{a\over a_0}\right)\right]
\,,  \label{rW_a}
\end{equation}
which describes Alfv\'{e}n waves.
The Alfv\'{e}n frequency
\begin{equation}
\nu_{\rm a}={H_0\over\pi}\left({k\over k_{{\rm
a}0}}\right)^2={\textstyle{1\over2}}\delta\nu_{\rm ac}\,,
\label{osc}\end{equation}
where $\delta\nu_{\rm ac}=\nu_{\rm ac,mag}-\nu_{\rm ac}$ is the
excess magnetic acoustic frequency given in Eq. (\ref{3}).
Thus local differential vortices in the density distribution are
``flip-flopping" in concert with the acoustic oscillations in the
density perturbations.
Alfv\'{e}n waves are a purely magnetic effect, arising from
the fluctuations in the magnetic field direction.
These waves have decaying amplitude, in common with non-magnetized
(and non-propagating) vector perturbations.

On scales well beyond the Alfv\'{e}n horizon, $k\ll k_{{\rm a}0}$,
the oscillatory behavior
is not felt, and Eq. (\ref{rW_a}) gives, to lowest order,
\[
{\cal W}=C_{(1)}\left[{a_0\over
a}-{\textstyle{1\over2}}\left({k\over k_{{\rm a}0}}
\right)^2\left({a\over
a_0}\right)\right]+C_{(2)}
\left({ k\over k_{{\rm a}0}}\right)\,.
\]
On superhorizon scales, the oscillations
disappear: ${\cal W}\rightarrow C_{(1)}(a_0/a)$, regaining the
standard non-magnetized result.
Thus, before matter-radiation
equality, and on scales much larger than the Alfv\'{e}n horizon,
density vortices evolve unaffected by the presence of a
cosmological magnetic field.

After equality, in the matter-dominated dust era, Eq.
(\ref{nddotW_a}) becomes
\begin{equation}
3\left(\frac{t}{t_0}\right)^2{\cal W}''+8\left({t\over t_0}\right)
{\cal W}'+\left[2+
\left({k\over k_{{\rm a}0}}\right)^2\right]{\cal W}=0 \,,   \label{dnddW_a}
\end{equation}
where a prime denotes $d/d(t/t_0)$.
The solution is
\begin{equation}
{\cal W}=C_{(+)}\left({t\over t_0}\right)^{\alpha_+}+
 C_{(-)}\left({t\over t_0}\right)^{\alpha_-}\,, \label{dW_a}
\end{equation}
where
\begin{equation}
\alpha_\pm={\textstyle{1\over6}}\left[-5
\pm\sqrt{1-12 \left(
{k\over k_{{\rm a}0}}\right)^2}\,\right]
 \,.  \label{dz12}
\end{equation}
Therefore, any rotational instabilities present in the density
distribution of the dust die away with time, as they do in
non-magnetized cosmologies. The field effect on a given mode $k$ is
to reduce the depletion rate of $W_a$ by an amount proportional to
the initial Alfv\'{e}n speed squared, $(c_{\rm a}^2)_0$. Thus,
magnetized dust universes will contain more residual vortices than
magnetic-free ones. However, the effect is confined within a narrow
wavelength band beyond the Alfv\'{e}n horizon $\lambda_{\rm a}$. On
much larger scales, the field
influence becomes negligible, and $W_a\propto t_0/t$ as in
non-magnetized models.
On scales with $k>k_{{\rm a}0}/\sqrt{12}$, i.e. within a few times
the Alfv\'{e}n scale, Eq. (\ref{dz12}) shows that the density
vortices oscillate as Alfv\'{e}n waves.

%%%%%%%%%%%%%%%%%%%%%%%%%%%%%%%%%%%%%
\section{Magnetized shape-distortion}
%%%%%%%%%%%%%%%%%%%%%%%%%%%%%%%%%%%%%

We monitor anisotropic deformation (shape distortion)
 in the density distribution of the medium
through the PSTF tensor
$\xi_{ab}=a\D_{\langle a}\Delta_{b\rangle}$.
This is associated with density variations that do not
represent matter aggregations,
since the associated
divergence of $\Delta_a$ is zero, but rather describe changes in
the local anisotropy pattern of the density gradients.

Distortion in the density is coupled to
distortion in the expansion and the magnetic energy
density, defined via the PSTF tensors
\begin{equation}
\vartheta_{ab}=a^2\D_{\langle  a}\D_{b\rangle}\Theta\,,~~
\beta_{ab}=
\frac{a^2}{B^2}\D_{\langle  a}\D_{b\rangle}B^2 \,,  \label{vthe_ab-bet_ab}
\end{equation}
which vanish in the background and are thus gauge-invariant.
The propagation equations for $\xi_{ab}$, $\vartheta_{ab}$ and $\beta_{ab}$
follow from the comoving PSTF derivatives of Eqs.
(\ref{dotDel_a})--(\ref{dotB_ab}):
\begin{eqnarray}
\dot{\xi}_{ab}&=& 3wH\xi_{ab}- (1+w)\vartheta_{ab}+
{\textstyle{3\over2}}c_{\rm a}^2H\beta_{ab}- c_{\rm
a}^2H\kappa_{ab}- 3H\mu_{ab} \,,  \label{dotxi}\\
\dot{\vartheta}_{ab}&=&-2H\vartheta_{ab}-
{\textstyle{1\over2}}{\rho}\xi_{ab}- \frac{c_{\rm
s}^2}{1+w}\D^2\xi_{ab}+ {\textstyle{1\over4}}{c_{\rm
a}^2\rho}\beta_{ab}- \frac{c_{\rm
a}^2}{2(1+w)}\D^2\beta_{ab}\nonumber\\
&&{}-{\textstyle{1\over2}}{c_{\rm a}^2\rho}\kappa_{ab}-
{\textstyle{3\over2}}\rho\mu_{ab}- \left(6c_{\rm
s}^2+\frac{4c_{\rm a}^2}{1+w}\right)H\varpi_{ab} \,,
\label{dotvth}\\
\dot{\beta}_{ab}&=&\frac{4}{3(1+w)}\dot{\xi}_{ab}+ \frac{4(c_{\rm
s}^2-w)H}{1+w}\xi_{ab} \,.  \label{dotbet}
\end{eqnarray}
The additional
gauge-invariant PSTF tensors
\[%\begin{equation}
\kappa_{ab}=a^2{\cal R}_{\langle ab\rangle}\,,~~
\varpi_{ab}=a^2\D_{\langle a}\mbox{curl}\,\omega_{b\rangle}\,,~~
\mu_{ab}={a\over\rho}B^c\D_cB_{\langle  ab\rangle} \,,
\]%\label{vk,vp&mu}\end{equation}
respectively describe distortions caused by projected
curvature, rotation and by anisotropies in the distribution of the
magnetic field gradients. The first is due to the natural coupling
of the field to the curvature and is given by
\[%\begin{equation}
\kappa_{ab}=a^2\left({\textstyle{1\over2}}\pi_{ab}-
H\sigma_{ab}+
E_{ab}\right)\,,
\]%\label{kappa_ab}\end{equation}
obtained from Eq. (\ref{3R_ab}) by means of the shear propagation
equation (\ref{dotsh}).
The second arises from the fluid flow, which
generally is not hypersurface orthogonal. It has no impact
on deformation if the rather special condition
$\D_{\langle  a}\mbox{curl}\,\omega_{b\rangle}=0$ holds.
Finally, the effect of $\mu_{ab}$ vanishes when any
anisotropies present in the distribution of $B_{ab}\equiv a\D_bB_a$ remain
invariant along the magnetic force-lines, that is when
$B^c\D_cB_{\langle  ab\rangle}=0$.
Note that both the scalar and vector aspects of the field
contribute to shape distortion.

Equations (\ref{dotxi}) and (\ref{dotvth}) combine to provide a
second order differential equation, also obtained by
taking the comoving PSTF derivative of
Eq. (\ref{ddotDel_a}). With $\Lambda=0$, we
have
\begin{eqnarray}
\ddot{\xi}_{ab}&=&-\left(2+3c_{\rm s}^2-6w\right)H\dot{\xi}_{ab}+
{\textstyle{1\over2}}
\left(1-6c_{\rm s}^2+8w-3w^2\right)\rho\xi_{ab}+
c_{\rm s}^2\D^2\xi_{ab}
\nonumber\\&\mbox{}&-{\textstyle{1\over2}}
\left(1-3c_{\rm s}^2+2w\right)c_{\rm a}^2\rho\beta_{ab}+
{\textstyle{1\over2}}c_{\rm
a}^2\D^2\beta_{ab}+{\textstyle{1\over3}}
\left(2-3c_{\rm s}^2+3w\right)\left(c_{\rm a}^2\rho\kappa_{ab}+3\rho\mu_{ab}\right)
\nonumber\\&\mbox{}&+
\left[6(1+w)c_{\rm s}^2+2c_{\rm a}^2\right]H\varpi_{ab} \,.  \label{ddotxi_ab}
\end{eqnarray}
This is coupled to Eq. (\ref{dotbet}) for the growth of
infinitesimal distortions in the magnetic energy density. The
other source terms, namely $\kappa_{ab}$, $\mu_{ab}$ and
$\varpi_{ab}$, evolve according to
\begin{eqnarray}
\dot{\kappa}_{ab}&=&-\frac{1}{1+w}\dot{\xi}_{ab}+ \frac{(c_{\rm
s}^2+3w)H}{1+w}\xi_{ab}+ \frac{2c_{\rm a}^2H}{1+w}\beta_{ab}-
\frac{4H}{1+w}\mu_{ab}-2\varpi_{ab}- \D^2\sigma_{ab} \,,
\label{dotkap}\\ \dot{\mu}_{ab}&=&-\left[4-3(1+w)\right]H\mu_{ab}+
\frac{c_{\rm a}^2}{3(1+w)}\dot{\xi}_{ab}+ \frac{(c_{\rm
s}^2-w)c_{\rm a}^2H}{1+w}\xi_{ab} \nonumber\\ &&{}- \frac{c_{\rm
a}^4H}{2(1+w)}\beta_{ab}+ \frac{c_{\rm a}^4H}{3(1+w)}\kappa_{ab}+
{\textstyle{1\over3}}c_{\rm a}^2a^2\D^2\sigma_{ab} \,,
\label{dotmu}\\ \dot{\varpi}_{ab}&=&-(2-3c_{\rm s}^2)H\varpi_{ab}+
\frac{a}{2(1+w)\rho}\D_c\D_{\langle
a}B^d\D_{|d|}\left[B_{b\rangle}{}^c-B^c{}_{b\rangle}\right]\,.
\label{dotvpi}
\end{eqnarray}
We used
the propagation equations (\ref{gem1}) and (\ref{pi})
for $E_{ab}$ and $\pi_{ab}$, which
 imply
\[
\dot{E}_{ab}=-3HE_{ab}+
{\textstyle{3\over2}}H\pi_{ab}-
{\textstyle{1\over2}}(1+w)\rho\sigma_{ab}-
\D^2\sigma_{ab}+
\frac{1}{a^2}\left(\vartheta_{ab}+2\varpi_{ab}\right) \,,
%\label{dotE}
\]
on using the
constraint equations (\ref{shcon}) and (\ref{Hcon}).
Equation (\ref{dotmu}) requires Eqs.
(\ref{Max2}) and
(\ref{dotxi}). Finally, to obtain the evolution formula of
$\varpi_{ab}$ we have successively taken the comoving curl and
the comoving PSTF derivative of Eq. (\ref{dotom}).

The system of equations (\ref{dotbet})--(\ref{dotvpi}) provides in principle a
complete description of linear infinitesimal shape distortion generated
by magnetic effects, provided we have a prescription for the $\D^2\sigma_{ab}$
terms in Eqs. (\ref{dotkap}) and (\ref{dotmu}),
and for the last term on the right of Eq. (\ref{dotvpi}). Even
without these terms, the system is too complicated to
analyze in general. However,
it is clear in general terms how magnetic effects will actively
generate distortion. We can illustrate this
by comparing with the non-magnetized case in a simple example.

For simplicity, consider superhorizon scales in the dust era
(neglecting vorticity).
Suppose that at a given event $(t_0,\vec{x}_0)$, we have no
initial distortion or rate of distortion:
\begin{equation}
(\xi_{ab})_0=0=(\dot{\xi}_{ab})_0\,,
\label{d1}\end{equation}
In the non-magnetized case, the distortion system collapses
to the single equation
\[
\ddot{\xi}_{ab}=-2H\dot{\xi}_{ab}+
{\textstyle{3\over2}}H^2\xi_{ab}\,.
\]
It follows that along the
fluid flow line through $\vec{x}_0$, no distortion is generated:
\[
\xi_{ab}(t,\vec{x}_0)=0\,.
\]
The evolution is purely {\em passive}, or inertial,
i.e., distortion can only develop if it is there {\em a priori}.
In the magnetized case, by contrast, distortion is {\em actively} and
nonadiabatically generated by magnetic effects.
Equation (\ref{ddotxi_ab}) shows that
\[
H_0^{-2}(\ddot{\xi}_{ab})_0=\left(c_{\rm
a}^2\right)_0\left[-{\textstyle{3\over2}}\left(\kappa_{ab}\right)_0+
2\left(\kappa_{ab}\right)_0\right]+6\left(\mu_{ab}\right)_0\,,
\]
so that $(\ddot{\xi}_{ab})_0\neq0$. Distortion is immediately
generated along the flow line. In fact, the distortion has a
growing mode, as we now show.

The superhorizon scalar modes of the distortion
tensors satisfy a system that follows from Eqs. (\ref{dotbet}) and
(\ref{ddotxi_ab})--(\ref{dotmu}):
\begin{eqnarray}
\xi''&=&-{\textstyle{4\over3}}\left({t_0\over t}\right)\xi'+
{\textstyle{2\over3}}\left({t_0\over t}\right)^2\xi-
{\textstyle{2\over9}}\left(c_{\rm a}^2\right)_0\left({t_0\over t}\right)^{8/3}
[3\beta-4\kappa]+
{\textstyle{8\over3}}\left({t_0\over t}\right)^2\mu \,,  \label{dnddotxi}\\
\beta'&=&{\textstyle{4\over3}}\xi' \,,  \label{dndotbet}\\
\kappa'&=&-\xi'+{\textstyle{4\over3}}\left(c_{\rm a}^2\right)_0
\left({t_0\over t}\right)^{5/3}\beta-
{\textstyle{8\over3}}\left({t_0\over t}\right)\mu \,, \label{dndotkap}\\
\mu'&=&-{\textstyle{2\over3}}\left({t_0\over
t}\right)\mu+{\textstyle{1\over3}}\left(c_{\rm
a}^2\right)_0\left({t_0\over t}\right)^{2/3}\xi'\,.
\label{dndotmu}
\end{eqnarray}
Equations (\ref{dndotbet}) and (\ref{dndotmu}) integrate to
\begin{equation}
\beta={\textstyle{4\over3}}\left(\xi+\Gamma_\beta\right)\,,~~
\mu={\textstyle{4\over3}}\left(c_{\rm a}^2\right)_0
\left({t_0\over t}\right)^{2/3}
\left[\xi+\Gamma_\mu\right] \,,  \label{dnbetmu}
\end{equation}
where $\dot{\Gamma}_\beta=0= \dot{\Gamma}_\mu$. Then
Eq. (\ref{dnbetmu}) transforms Eq. (\ref{dndotkap}) into
\begin{equation}
\kappa'=-\xi'+{\textstyle{8\over9}}\left(c_{\rm a}^2\right)_0
\left({t_0\over
t}\right)^{2/3}\left[\xi+2\Gamma_\beta-\Gamma_\mu\right]\,.
\label{dndotkap1}
\end{equation}
According to Eq. (\ref{dnbetmu}), the effect of any anisotropies
present in the distribution of $B_{ab}\equiv a\D_aB_b$ on $\xi$
decreases after matter-radiation equilibrium. Since $\mu$
is a key source of shape-distortion, we expect the evolution
of $\xi$ to approach that of $\Delta$ as the universe expands.
Equation (\ref{dnddotxi}) gives
\[
9\left({t\over t_0}\right)^3\xi'''+36\left({t\over t_0}\right)^2\xi''+
14\left({t\over t_0}\right)\xi'-
4\xi=0 \,,
\]
on using Eqs. (\ref{dnbetmu}) and (\ref{dndotkap1}).
This has the same form as the corresponding density perturbation
equation (\ref{ldndddotDel}), and the solution is thus of the form Eq.
(\ref{lldnDel}). Imposing the initial conditions in Eq. (\ref{d1}), we find
that
\begin{equation}
\xi(t,\vec{x}_0)=\Gamma\left[\left({t\over t_0}\right)^{2/3}+4\left({t\over t_0}
\right)^{-1}-5\left({t\over t_0}\right)^{-2/3}\right] \,,  \label{lldnxi}
\end{equation}
where $\Gamma$ is a constant.
Thus the shape distortion has a growing mode along the fluid flow
line, due purely to magnetic effects; in the absence of the
magnetic field, $\Gamma=0$.
(A similar situation arises in the simpler case of distortion generated by
velocity dispersion \cite{mtm}.)

%%%%%%%%%%%%%%%%%%%%
\section{Conclusion}
%%%%%%%%%%%%%%%%%%%%

We have given a fully general relativistic treatment of
the scalar and vector effects of a weak large-scale magnetic field
on cosmological density inhomogeneity. This
refines the results of \cite{TB} on magnetized
density perturbations, and extends that work to
analyze magnetized vortices and
shape distortion in the density distribution.
Our covariant Lagrangian approach allows us to
derive gauge-invariant
evolution equations for all these aspects of density
inhomogeneity in the general linear case, i.e., incorporating
all fluctuations of, and couplings between, the field, the fluid and the
curvature. In summary, magnetized density perturbations are governed by
Eqs. (\ref{ddotDel})--(\ref{dotcK}), magnetized density vortices are governed
by Eq. (\ref{ddotW_a}), and magnetized shape distortion is
governed by Eqs. (\ref{dotbet})--(\ref{dotvpi}).

We give the solutions in closed form for magnetized density perturbations and
vortices, in the radiation and dust eras. Some of the scalar
solutions and all of the vector solutions are new.
For magnetized shape
distortion, we found a special solution with a growing mode.

Given the overall weakness of the field, the magnetic
effects described here are secondary relative to those of the
matter. In some
cases, second-order fluid effects may be comparable in strength
to first-order magnetic effects.
However, their presence does
not affect the field impact, which remains the same.
Thus we can neglect the second-order effects in a consistent
linear analysis that probes the lowest order magnetic effect on
density inhomogeneity.
To lowest
order, the magnetic influence on gravitational instability may be
summarised as follows:

\begin{enumerate}
\item
The magneto-curvature coupling, which is a direct consequence of
the field's vectorial nature, first identified in \cite{TB}, has
an important influence. In particular, on superhorizon scales
in the radiation era, this coupling slightly enhances the growing
mode of density perturbations, as shown in Eqs.
(\ref{lrnDel}) and (\ref{nc}):
\[
\Delta= C_{(+)}\left({a\over a_0}\right)^{2+{1\over2}c_{\rm a}^4}
+C_{(1-)}\left({a\over a_0}\right)^{-1+c_{\rm a}^2}+C_{(0)}+
C_{(2-)}\left({a\over
a_0}\right)^{-c_{\rm a}^2}\,.
\]
When the coupling is neglected, the
growing mode is incorrectly found to be damped relative to the non-magnetized
case.

\item
Magneto-sonic waves in the radiation era are given in exact form in Eq.
(\ref{srnDel}),
\begin{eqnarray*}
\Delta&=&\left[C_{(1)}-C_{{\cal K}}
{\rm Si}\left(\beta {k\over k_{{\rm h}0}}{a\over a_0}\right)
\right]\sin\left(\beta {k\over k_{{\rm h}0}}{a\over a_0}\right)\nonumber\\
&&{}+\left[C_{(2)}-C_{{\cal K}}
{\rm Ci}\left(\beta {k\over k_{{\rm h}0}}{a\over a_0}\right)
\right]\cos\left(\beta {k\over k_{{\rm h}0}}{a\over a_0}\right)
-C_{{\cal B}}c_{\rm a}^2\,.
\end{eqnarray*}
This shows the nonadiabatic modulation of the amplitude and
increase in the frequency of acoustic oscillations. These effects,
together with the nonzero average value implied by the $C_{{\cal B}}$
term, have
potentially important implications for the CMB acoustic peaks,
some of which have been investigated in \cite{adgr}.

\item
In the dust era, subhorizon magnetized density perturbations are
given exactly in Eq. (\ref{sdnDel}), which leads to the magnetized
Jeans scale in Eq. (\ref{mj}):
\[
\lambda_{\rm mJ}(t_0)={\textstyle{8\over15}}\pi\sqrt{6}
\left(c_{\rm a}\right)_0\lambda_{{\rm h}0}\,.
\]
On scales such that $\lambda_{\rm mJ}\ll\lambda\ll\lambda_{\rm
h}$, the density perturbations are
\[
\Delta=C_{(+)}\left({t\over t_0}\right)^{{2\over3}-\epsilon}+
C_{(-)}\left({t\over t_0}\right)^{-1+\epsilon}+
C_{({\cal B}-)}\left({t\over t_0}\right)^{-2/3}
-{\textstyle{5\over2}}\epsilon
C_{\cal B}\,,
\]
where $\epsilon={2\over5}(k/k_{{\rm a}0})^2$. This shows the small
damping effect on the adiabatic growing mode, as well as the new
nonadiabatic modes. These results imply small modifications to
structure formation in the linear regime. However, they are
limited by the fact that we have neglected any non-baryonic matter
or cosmological term.

\item
Pure-magnetic density fluctuations, which are induced in an
initially smooth fluid by magnetogenesis, are given on
superhorizon scales by Eq. (\ref{new'}). This solution would be
important in any attempt to model
large-scale structure formation as seeded by magnetogenesis.

\item
Magnetized isocurvature perturbations are characterized by Eq.
(\ref{iso1}). On superhorizon scales, these modes are purely
decaying.

\item
The field is a source of incompressible rotational
instabilities, and the condition for this to happen is given via
Eq. (\ref{dotW_a}).
Magnetized density vortices are shown be scale-dependent
and to precess, at a rate given
by Eq. (\ref{prec}).

The general propagation equation for these vortices
(i.e., incorporating all relevant effects) is given by Eq.
(\ref{ddotW_a}):
\[
\ddot{W}_a+\left(4-3w\right)H\dot{W}_a+
{\textstyle{1\over2}}\left[1-7w+3c_{\rm s}^2(1+w)\right]\rho W_a=
\left[\frac{c_{\rm a}^2}{3(1+w)}\right]\D^2W_a \,.
\]
In the radiation era, the Alfv\'{e}n wave solutions are given exactly
in Eq. (\ref{rW_a}). The Alfv\'{e}n frequency and wave-speed are
\[
\nu_{\rm a}={H_0\over\pi}\left({k\over k_{{\rm
a}0}}\right)^2\,,~~
  v_{\rm a}={\textstyle{1\over2}}c_{\rm a}\,.
\]
These results generalize some of the theoretical results of Durrer
et al.
\cite{dky}, who then go further and apply the results to determine the effect
of Alfv\'{e}n wave modes on CMB anisotropies.

\item
After recombination, magnetized density vortices are given exactly in
 Eq. (\ref{dW_a}). They decay like their
adiabatic counterparts, but at a slower rate, so that rotational
instability persists for longer in a magnetic universe.
This will have a small effect on structure formation in the linear
regime.

\item
Finally, we have investigated for the first time
magnetic effects on infinitesimal shape distortion in the
density distribution. The magnetic influence is
manifold. Anisotropies in the field energy density,
together with those in the distribution of the magnetic vector
itself are direct sources of density deformation. The field's
coupling to curvature and rotation also acts as an indirect source
of magnetically induced shape distortions. Following the evolution
of shape-distortion along the worldline of a fluid
element, we showed that the field is an {\em active} source of
distortion. On superhorizon scales, we showed
via a special solution of the shape-distortion system that
there is a growing mode of magnetized shape-distortion [see
 Eq. (\ref{lldnxi})].

Unlike the magnetic effects on density and rotational
perturbations, which are small corrections of the non-magnetized
results, magnetic effects on shape-distortion constitute a
significant change from the non-magnetized (and {\em passive})
case. (A similar statement applies in the case of velocity
dispersion effects in CDM \cite{mtm}.)
The results on magnetized shape-distortion have potentially
important implications for (linear) structure formation. Not only
is distortion actively generated once scales re-enter the Hubble
horizon and begin to collapse, but it is also actively generated
while the scales are beyond the horizon. Of course, the shape
distortion in the linear regime will be overwhelmed by effects
that arise during the nonlinear stages of collapse.

\end{enumerate}

\[ \]
{\bf Acknowledgments:} CGT is supported by PPARC. We thank Marco Bruni
and David Matravers for useful discussions.

\end{document}